\documentclass{article}
\usepackage{arxiv}

\usepackage[utf8]{inputenc} % allow utf-8 input
\usepackage[T1]{fontenc}    % use 8-bit T1 fonts
\usepackage{hyperref}       % hyperlinks
\usepackage{url}            % simple URL typesetting
\usepackage{booktabs}       % professional-quality tables
\usepackage{amsfonts}       % blackboard math symbols
\usepackage{amsmath}
\usepackage{nicefrac}       % compact symbols for 1/2, etc.
\usepackage{microtype}      % microtypography
\usepackage{graphicx}
\usepackage{natbib}
\usepackage{doi}

\title{Development and Stability Analysis of Carpal Kinematic Metrics from 4D Magnetic Resonance Imaging}

\author{ \href{https://orcid.org/0000-0001-9572-4922}{\includegraphics[scale=0.06]{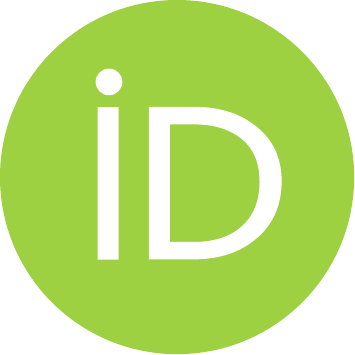}\hspace{1mm}Azadeh ~Sharafi} \\
	Center for Imaging Research\\
	Department of Radiology\\
	Medical College of Wisconsin  \\
	\texttt{asharafi@mcw.edu} \\
        \AND
	  {\hspace{1mm}Andrew S. ~Nencka} \\
	Center for Imaging Research\\
	Department of Radiology\\
	Medical College of Wisconsin  \\
	\texttt{anencka@mcw.edu} \\
	\And
	{\hspace{1mm}Kevin M.~Koch}\\
	Center for Imaging Research\\
	Department of Radiology\\
	Medical College of Wisconsin  \\
	\texttt{kkoch@mcw.edu} \\
}

\renewcommand{\shorttitle}{Kinematic Metrics from 4D MRI}

\hypersetup{
pdftitle={Wrist Kinematics progress report},
pdfsubject={q-bio.NC, q-bio.QM},
pdfauthor={Azadeh ~Sharafi, Andrew S. Nencka, Kevin K.~Koch},
pdfkeywords={First keyword, Second keyword, More},
}

\begin{document}
\maketitle

\begin{abstract}
\noindent
\textbf{Introduction:} \\
Wrist instability remains a common health concern. The potential of dynamic Magnetic Resonance Imaging (MRI) in understanding and assessing carpal dynamics associated with this condition is a field of ongoing research. This study contributes to this line of inquiry by developing MRI-derived carpal kinematic metrics and investigating their stability.

\noindent
\textbf{Methods:} \\
A previously described 4D MRI approach for tracking the movements of scaphoid, lunate, and capitate bones in the wrist was deployed in this study. A panel of 120 metrics characterizing radial/ulnar deviation and flexion extension movements was constructed by fitting low order polynomial models of scaphoid and lunate degrees of freedom against that of the capitate. Intraclass Correlation Coefficients (ICC) were utilized to analyze intra- and inter-subject stability within a mixed cohort of 49 subjects, including 20 with a history of wrist injury and 29 without such history.

\noindent
\textbf{Results:} \\
A comparable degree of stability across the two different wrist movements. Out of the total 120 derived metrics, distinct subsets demonstrated high stability within each type of movement. Among asymptomatic subjects, 16 out of 17 metrics with high intra-subject stability also showed high inter-subject stability. Interestingly, some quadratic term metrics, although relatively unstable within asymptomatic subjects, showed increased stability within this group, hinting at potential differentiation in their behavior across different cohorts.

\noindent
\textbf{Conclusion:} \\
The findings of this study demonstrate the developing potential of dynamic MRI to assess and characterize complex carpal bone dynamics.  
Stability analyses of the derived kinematic metrics showed encouraging differences between cohorts with and without a history of wrist injury.   
Although these broad metric stability variations highlight the potential utility of this approach for analysis of carpal instability, further studies are necessary to better characterize these observations.
\end{abstract}

% keywords can be removed

\section{Introduction}
Carpal instability stands as a significant health issue with the potential to drastically impair individuals' quality of life. It has been reported that perilunate injuries account for roughly 7$\%$ to 10$\%$ of all wrist injuries \cite{melsom2007carpal, muppavarapu2015perilunate}, and they make up 19$\%$ of sprains that occur without concomitant fractures \cite{jones1988beware}. Distal radius fractures appear to have an even stronger correlation with carpal instability, as the incidence rate of the latter can be as high as 30$\%$ in those experiencing the former \cite{tang1992carpal}. A recent study by O'Brien et al.\cite{OBRIEN2018282} provides an even more compelling statistic, reporting a cumulative incidence rate of 44$\%$ for carpal instability within two years following an injury.

Carpal instability, marked predominantly by wrist discomfort and diminished functional capacity, can greatly restrict an individual's ability to execute everyday tasks. This condition can interfere with basic actions such as holding a cup, as well as more intricate tasks such as keyboard typing or playing a musical instrument. Carpal instability is a complex and varied pathological condition of the wrist. It may result from the carpus' inability to withstand physiological stressors, or due to abnormal kinematics of the carpal bones during activity \cite{2016Ramamurthy, schmitt2006carpal}. The term "dynamic" instability refers to a deformity that only becomes apparent during motion, as opposed to "static" instability, which is observable even when the wrist is at rest \cite{schmitt2006carpal}.

The conventional approach to diagnosing and studying carpal instability has predominantly hinged on static imaging modalities, such as conventional radiography, computed tomography (CT), or magnetic resonance imaging (MRI). These static imaging techniques have proven indispensable for investigating the wrist's structural integrity and alignment under static conditions \cite{2016Ramamurthy}. The key strengths of static imaging modalities reside in their proven effectiveness in identifying fractures, evaluating bone alignment, ligament damage, and observing arthritic changes. However, their inherent limitation is that they capture only a static moment in time, thus failing to elucidate the dynamic interplay and kinematics of the carpal bones. This lack of dynamic context may lead to an incomplete or skewed interpretation of the severity and traits of the instability.

The quest for a deeper understanding of wrist kinematics in both health and disease has sparked burgeoning interest in the application of dynamic imaging techniques to delineate the kinematics of carpal bones and examine their interrelationships during wrist motion \cite{choi2013four, berdia2006hysteresis, toms2011midcarpal, crisco1999noninvasive, brinkhorst2021quantifying}. Fluoroscopy has been utilized to observe dynamic joint movement, especially when movement primarily rotates about a single axis, and for diagnosing severe instabilities or abnormalities \cite{tashman2003vivo}. However, fluoroscopy's inherent two-dimensional nature limits its ability to detect intricate and subtle musculoskeletal discrepancies, such as instabilities in the wrist joint \cite{tashman2003vivo}.

Dynamic CT and MRI, on the other hand, generate a sequence of time-resolved three-dimensional images, enabling a non-invasive investigation into the individual kinematics of wrist bones. Dynamic CT, in particular, can provide high-resolution three-dimensional images of the wrist in motion, assisting in the visualization of the relative motion of carpal bones \cite{choi2013four}. The principal drawback of dynamic CT, however, is its use of ionizing radiation, which imposes restrictions on its use in specific patient populations and in longitudinal studies.

MRI, on the other hand, offers volumetric imaging with excellent soft tissue contrast without the use of ionizing radiation. This makes MRI it an attractive option for dynamic evaluation of the wrist \cite{2016Ramamurthy,Vassa2020}.   The combined challenges of acquisition inefficiency and vulnerability to motion artifacts have historically limited MRI's capability for dynamic analysis of the wrist.   Recent demonstrations using dynamic MRI have performed manual assessment and measurements of carpal bones during motion cycles~\cite{abbas2019analysis,Shaw_2019,henrichon2020dynamic}.  In addition, the work presented by Zarenia et al~\cite{zarenia2022dynamic} demonstrated that advanced clinical MRI sequences used to collect 4D datasets that can facilitate continuous and semi-automated carpal motion analysis during simple wrist motions.   

The present study seeks to build upon the methods of Zarenia et al.~\cite{zarenia2022dynamic}, focusing specifically on the creation of self-normalized metrics to gauge the relative motion of the scaphoid and lunate during radial-ulnar deviation and flexion-extension movements of the wrist.    We then analyze the stability of these metrics within repeated measures of a single subjects.   Finally, the stability of the derived metrics is evaluated across unique cohorts of healthy asymptomatic subjects and subjects with a history of wrist injury.

\section{Materials and methods}
\subsection{Evaluation Cohort}
% The study was approved by the Institutional Review Board (IRB) of the Medical College of Wisconsin. 28 participants (16 females, mean age 31$\pm$12 years; 12 males, mean age 33$\pm$10 years) with no prior wrist injuries or symptoms were recruited for these studies. Each participant was provided with written informed consent prior to their inclusion in the research. 

Evaluation of the proposed methodology  was carried out on a cohort comprising 49 subjects. The cohort was categorized into two groups based on their history related to wrist injuries.

The first group consisted of 29 individuals who had no previous history of wrist injuries or symptoms. This group had a female-to-male ratio of 16:13  of ages 32.0 $\pm$ 9.5 years. This group served as a control group, providing a baseline for comparison with the group with a history of wrist injuries.

The second analysis group comprised 20 subjects with a history of wrist injuries, with a female-to-male ratio of 12:8 of ages 41.7 $\pm$ 15.0 years. 

All subjects provided informed consent into an approved IRB protocol and all procedures were carried out in accordance with relevant local guidelines and regulations.

\subsection{Imaging}

Magnetic resonance imaging data were acquired using a GE Healthcare Signa Premier 3T MRI scanner, equipped with a 16-channel large flex coil and a 16-channel transmit/receive hand wrist coil. For the scanning procedure, participants were oriented in a head-first prone position. Only the wrist of the dominant hand of each subject was scanned.
The Liver Acquisition with Volume Acceleration (LAVA Flex, GE Healthcare) pulse sequence \cite{LAVAFlex} was employed to acquire high-resolution 3D static and low-resolution 4D dynamic imaging data. The pure water reconstructed image was used in this study.

\subsubsection{Static MRI protocol}
 Subjects were positioned in a supine position for collection of static images.   Images were acquired using the 16-channel QED transmit/receive hand wrist coil with the wrist/hand placed in the neutral position. Utilized imaging parameters were: voxel size = 0.9mm$\times$0.9mm$\times$1mm, matrix size = 256$\times$256$\times$60, TE = 1.7ms, TR = 5.3ms, flip angle=10$^{\circ}$, and bandwidth = 417 Hz/pixel, acquisition time = 2:21 min

\subsubsection{Dynamic MRI protocol}
For collection of dynamic imaging data, subjects were placed prone in the magnet bore with an extended wrist placed at the center of a 16-channel Flex array coil. Participants were instructed to perform two specific movements: radial-ulnar deviation and flexion-extension, commencing from the maximum ulnar or flexion position, respectively. These movements were repeated three times, guided by a visual cue presented during the MRI acquisitions.

While the participants had the liberty to perform these movements unrestricted, we ensured that the radius remained relatively stable through strategic padding of the subject's arm. For the radial-ulnar deviation motion, movements were executed on a flat surface to promote accuracy and consistency.

During each motion, 4D scans were obtained, comprising 40 LAVA Flex 3D volumes with a temporal spacing of 2.6 seconds.  The imaging parameters for the dynamic acquisition were: voxel size = 1.6mm$\times$1.6mm$\times$2.5mm, matrix size=128$\times$128$\times$12, compressed sensing  factor(CS) = 1.4, acceleration factor (R) = 2, TE = 1.2ms, TR=4ms, flip angle= 10$^{\circ}$, and bandwidth = 977 Hz/pixel, acquisition time = 2.5s

\begin{figure}
	\centering
	\includegraphics[width=8cm]{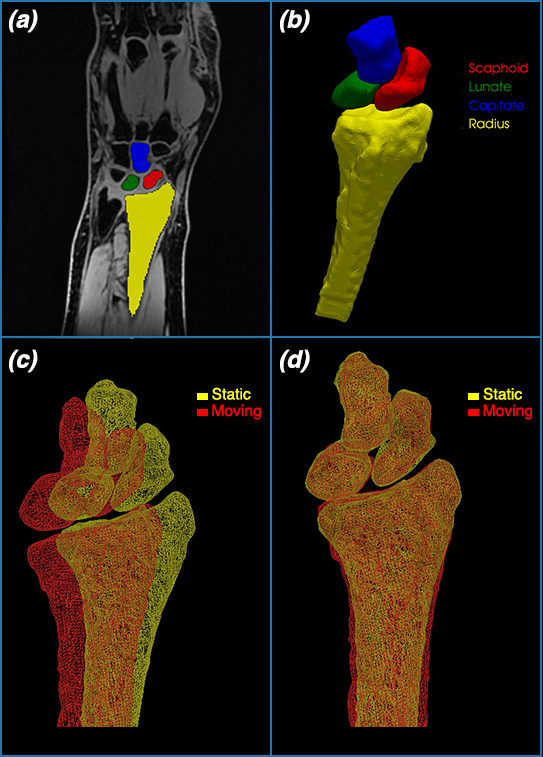}
	\caption{(a). Representative example of LAVAflex image and bone segmentation (b) Constructing a 3D surface mesh from 2D segmentations (c) Representative example of a static (yellow) and one moving frame (red) before registration (d) Registering the static and moving frame}
	\label{fig:fig1_2d_mesh}
\end{figure}

\subsection{Image Post-Processing}

\subsubsection*{Deep-Learning Based Automated Carpal Bone Segmentation} 
For the purposed of the present study, a deep convolutional neural network (CNN) was trained to automated segmentation of each bone separately.

\noindent \emph{Source training data:} 
The utilized neural network training approach relied heavily on augmentation of seed datasets curated by manual segmentation.     

The ITK-SNAP software \cite{itksnap} was used for manual segmentation of the scaphoid, lunate, capitate, and radius on a collection of images obtained from five volunteers (see Figure \ref{fig:fig1_2d_mesh}). Each dataset included 60 images for static and 960 images (480 per motion) for dynamic scans.These are the five datasets utilized in the preliminary analysis presented in Zarenia et al~\cite{zarenia2022dynamic}

 \subsubsection*{Network Training}  The network was implemented on NVIDIA Clara (v4.0.1) framework using MONAI (v0.8) \cite{MONAI2022} and trained on a workstation equipped with a 24-Intel(R) Core(TM) i9-10920X CPU @ 3.50GHz CPU and three GPUs (NVIDIA GeForce RTX 3090 with 24 GB of memory). 
 In our study, we employed an extensive data augmentation pipeline to artificially increase the diversity of our training set and mitigate overfitting. This was achieved by applying a sequence of transformations to each 3D image and its corresponding label in our dataset. Initially, the images and labels were loaded from disk and restructured to ensure the channel dimension was the first. Subsequently, a random spatial crop of size 128x128x8 was applied. The images and labels were then randomly flipped with probability of 0.5 along two axes, and the image intensities were randomly scaled and shifted. After normalizing the image intensities, random affine and 3D elastic deformations were applied to simulate the effect of natural variations in the data. Finally, the images and labels were converted into PyTorch tensors. By doing so, each image fed into the model during training was potentially unique due to the random transformations, effectively increasing the diversity of the training data and enhancing the model's generalization ability.
 
 SegResNet network \cite{MONAI2022} used, with  1, 2, 2, and 4 blocks in the downsampling path, and 1 block in each of the three stages of the upsampling path. The initial number of filters was set to 16. 
 A dropout probability of 0.2 was used to prevent overfitting.

The segmentation model was trained using the Adam optimizer with a learning rate of 1e-4 and weight decay of 1e-5. The learning rate was scheduled to follow a cosine annealing schedule over the course of 100 epochs. The DiceLoss function was used as the loss function. The DiceLoss was configured to exclude the background class, smooth the denominator by a small constant for numerical stability, square the predictions before computing the loss, convert the labels to one-hot encoding, and apply a softmax function to the predictions. Automatic mixed precision (AMP) was utilized during training to increase speed and efficiency. 

The model was validated periodically during training after each epoch. Inference was performed using a sliding window approach with an ROI size of [128,128,8] and an overlap of 0.5. The predictions were converted to a binary classification output by applying a softmax function and taking the argmax across the channel dimension.
The MeanDice metric, which calculates the Dice coefficient between the predictions and the labels, was used as the main performance metric during both training and validation.

\subsubsection*{Manual Segmentation Quality Inspection/Correction and Mesh Generation} 

A subset of segmentations was randomly selected and inspected in ITK-SNAP. Any imperfections found during the inspection were corrected. Afterward, the outer 3D surface mesh of each bone was reconstructed from binary labels using the well-known marching cube algorithm \cite{lorensen1987marching}.

\subsection{Radius-based registration}
The initial position and alignment of the processed models may exhibit considerable variation due to factors such as the inherent coordinate system of the MRI scanner and the potential movement of participants during data acquisition. To address this, we initially transformed the radius in all dynamic and static frames for each participant, aligning its first principal axis with the x-axis and orienting the +x towards the proximal direction. Following this, the Iterative Closest Point (ICP) algorithm \cite{icp} was employed to register all dynamic frames to their static counterparts. Identical transformations were subsequently applied to the remaining carpal bones.

\begin{figure}
	\centering
	\includegraphics[width=8cm]{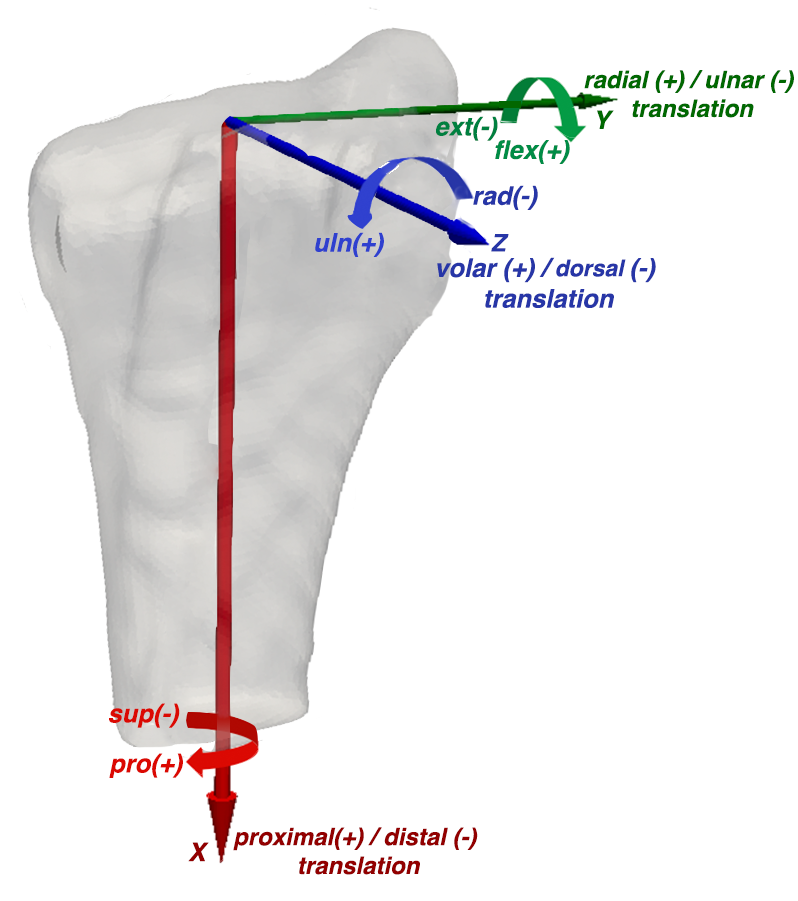}
	\caption{Radial coordinate system}
	\label{fig:fig2_rcs}
\end{figure}

The radial coordinate system is adapted from the recommendation of the International Society of Biomechanics
(ISB)\cite{wu2005isb}. The orientation of the X-axis (SP) was determined by aligning it with the central axis of the distal portion of the radius shaft (+X is proximal). Pronation (+) /supination(-) angles are defined around this axis. The Y-axis (RU) passes through the radial styloid peak and is orthogonal to the x-axis (+Y points radially). flexion(+)/extension (-) angles are defined around this axis. The Z-axis (RU) is the cross production of the x- and y-axis (+z is volar). The ulnar (+)/radial (-) deviation angles are defined around this axis.(Figure \ref{fig:fig2_rcs})

\subsection{Carpal Bone Motion Tracking}

In our approach to independently track the carpal bones during wrist movements, the degrees of freedom for computed carpal kinematics were anchored to a neutral wrist position, employing the radial coordinate system (RCS) as a reference point. 
\subsubsection*{Rigid Registration} 
The Iterative Closest Point (ICP) algorithm was used to align each wrist carpal bone from dynamic frames to their corresponding positions in the static frame. This rigis registration was performed individually for each bone.
A subset of the dataset was visualized using the pyvista package \cite{sullivan2019pyvista} to inspect the registration.

\subsubsection*{Outlier Frame Detection} Due to the inherent compromises between spatial resolution, temporal resolution, and signal-to-noise ratio of the acquired dynamic MR images, only a portion of the bone volume could be accurately imaged within each frame of the dynamic scans. This limitation can lead to poor carpal registrations in certain frames.

To mitigate this issue, we incorporated a strategy to exclude points that suffered from poor registration. This was achieved by applying a modification of the z-score, as proposed by Iglewicz and Hoaglin~\cite{iglewicz1993volume}. This adjusted z-score was computed based on the sum of the radius ($c_{rad}$) and the registration cost associated with the carpal bone ($c_{carpal}$).

\begin{equation}
\text{Modified } z = 0.675 \cdot \left|\frac{c_i - \tilde{c}}{MAD} \right|
\end{equation}

where $\tilde{c}$ is the median of  the registration cost $c = c_{rad} + c_{carpal}$, $c_i$ is the cost per time frame and MAD is the mean absolute deviation defined as : 
\begin{equation}
MAD = median_i\{\left|c_i - \tilde{c}\right| \}
\end{equation}

\subsubsection*{Calculate the metrics in RCS} 
After removing the outliers, the 4$\times$4 transformation matrices of each carpal bones for remaining frames were decomposed into a 3$\times$3 rotation matrix and a 3$\times$1 translation matrix. The rotation matrix was further converted into a set of three x-y-x Euler angles. This process effectively characterized the motion of the carpal bones as a combination of rotations about, and translations along, the three axes of the RCS. Translations were characterized by the displacement of each bone's center of mass within the RCS. This displacement was measured in three directions: distal-proximal (DP),  radial-ulnar (RU),and volar-dorsal (VD) \cite{akhbari2019}.

\subsubsection*{Independent Motion Cycle Segmentation}
 The  three repeated motions, or 'paths', taken during each subject's motion acquisition was automatically identified by pinpointing the local extrema in the capitate angle profile around the primary axis of movement. For the radial-ulnar deviation motion, this primary axis was the Z-axis, whereas for the flexion-extension motion, the primary axis was the Y-axis

The algorithm processes a time-series dataset and identifies points representing the motion cycle. It does so by identifying local maximums and minimums (extrema) within the data, facilitated by the argrelextrema function from Python's scipy.signal. The chosen extrema are then compiled into an array, representing the start, peak, trough, and end of the motion cycle.

In this process, the algorithm ensures the inclusion of the first and the last data points, modifying or appending to the extrema array as necessary. If more than four extrema points are identified, the algorithm enters an iterative process, selectively removing the least significant extrema (those with the smallest distance to their neighbors) until only four remain.

\begin{figure}
	\centering
	\includegraphics[width=15cm]{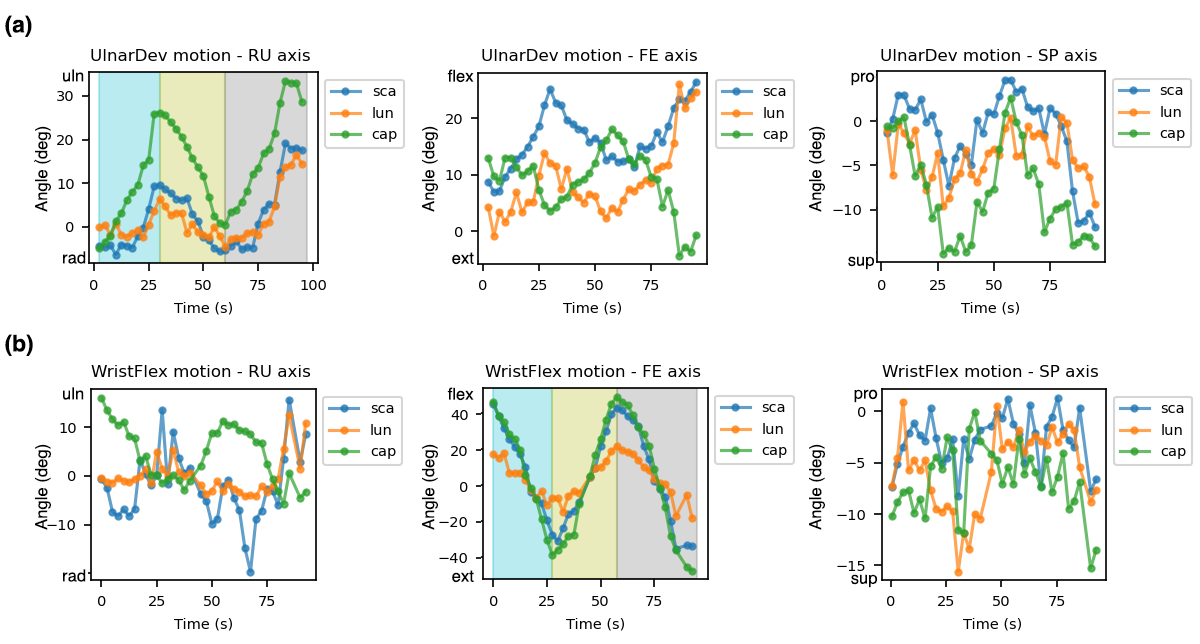}
	\caption{A representative illustration of the carpal bone's angular displacement in the RCS is provided for two types of motion: (a) radial-ulnar deviation, and (b) flexion-extension. The segmented paths are distinctly highlighted using three different background colors in the primary axis of movement for each motion. }
	\label{fig:fig3_time}
\end{figure}

The algorithm then outputs these four key indices within the input data which represent a unique motion cycle. It's important to note that the accuracy and efficiency of this method are contingent on the nature of the input data, particularly its periodicity and noise levels (Figure \ref{fig:fig3_time}).

% linear and a quadratic model is subsequently fitted to each path.    
%% XXXXXXX need the sample plots in here -- showing the path designations....

\subsection{Kinematic Metric Derivation}

It is known that the capitate and third metacarpal bone move synchronously throughout basic wrist motions~\cite{NEU2001_capRef}. Therefore, plotting scaphoid and lunate motion against the capitate offers a simple mechanism for self-referenced analysis. This was the approach utilized in our preliminary published demonstration of our acquisition and tracking methodology \cite{zarenia2022dynamic}. 
For the purposes of consolidated metric derivation, it was observed that most motion parameters exhibited low-order polynomial behavior consistently across subjects. 

\begin{figure}
	\centering
	\includegraphics[width=15cm]{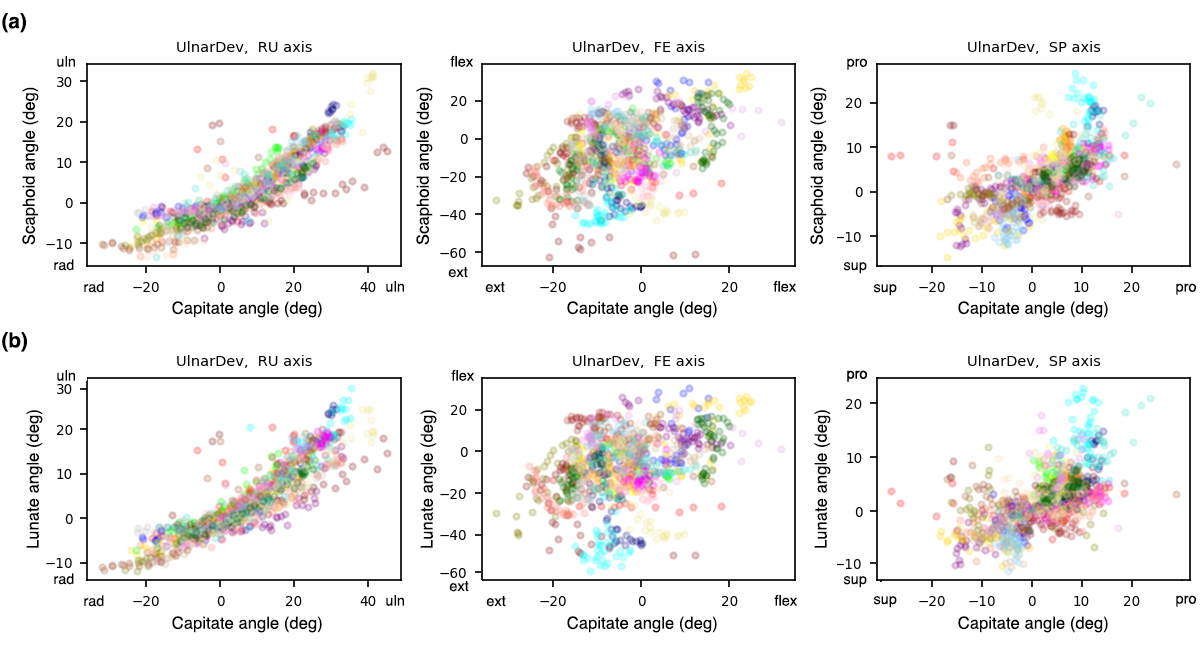}
	\caption{A representative illustration of (a) scaphoid and (b) lunate angles against capitate angle in the RCS during radial-ulnar deviation motion. }
	\label{fig:fig4_ulnardev_ang_cap}
\end{figure}

Figure \ref{fig:fig4_ulnardev_ang_cap} provides a a visualization of this plotting approach across the subset of present analysis cohort. Each subject in these plots is represented by a different color. The general low-order polynomial trends of the different metrics across the subject cohort are visible in these scatter plots. 

This observation was used to derive a battery of motion metrics across the motions, bones, and 6 kinematic parameters (i.e. rotations and translation ) 

Separate linear and quadratic fits were performed for each motion segment, here indicated as a “path”.   Regression was performed using the sklearn library to apply linear and quadratic fits to distinct motion paths within the input data. 

\begin{figure}
	\centering
	\includegraphics[width=13cm]{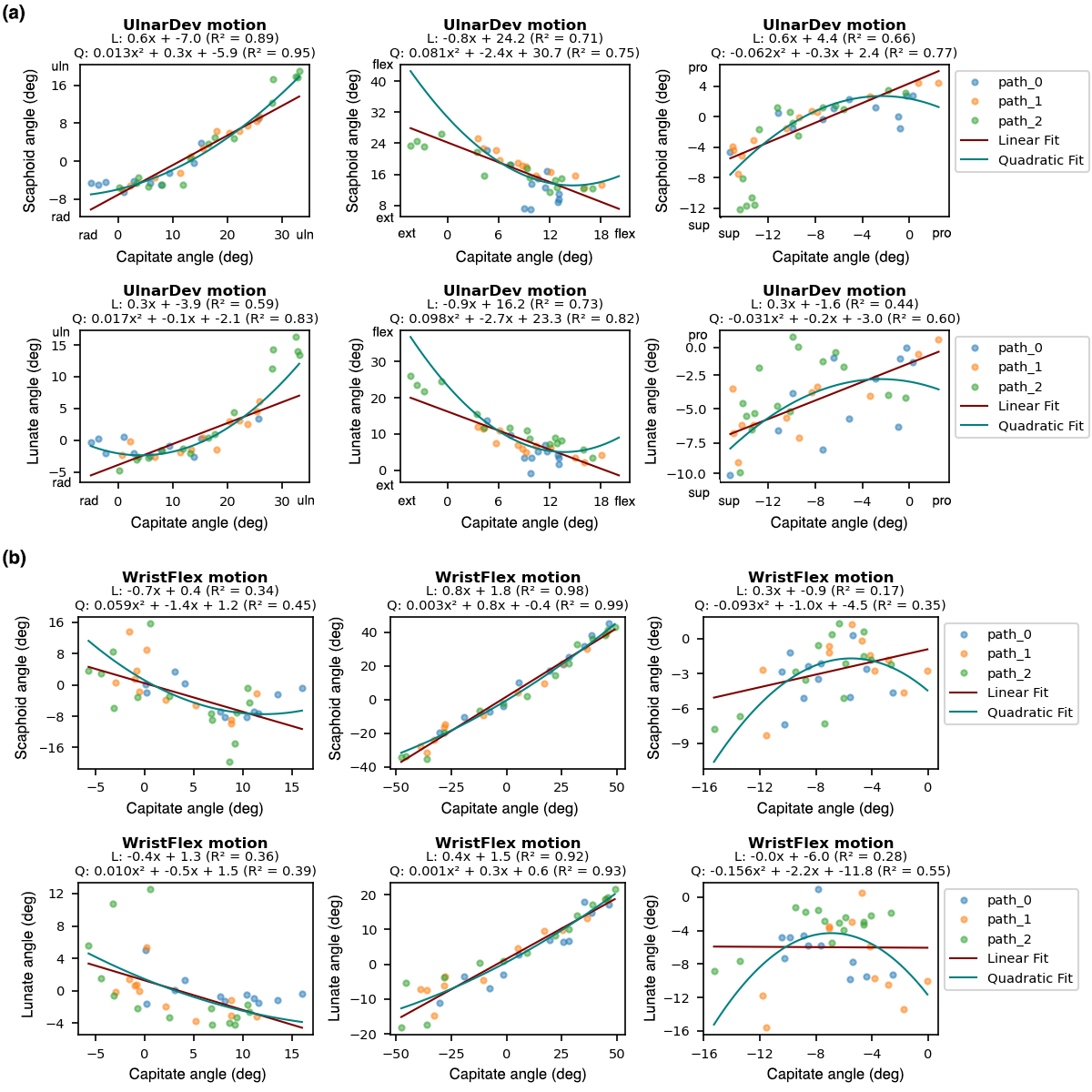}
	\caption{A representative depiction of linear and quadratic models are fitted to scaphoid and lunate angles versus the capitate angle, during (a) radial-ulnar deviation and (b) flexion-extension motion. The fits displayed in the figure represent the average fit calculated across three different paths. }
	\label{fig:fig5_ang_fit}
\end{figure}

\begin{figure}
	\centering
	\includegraphics[width=13cm]{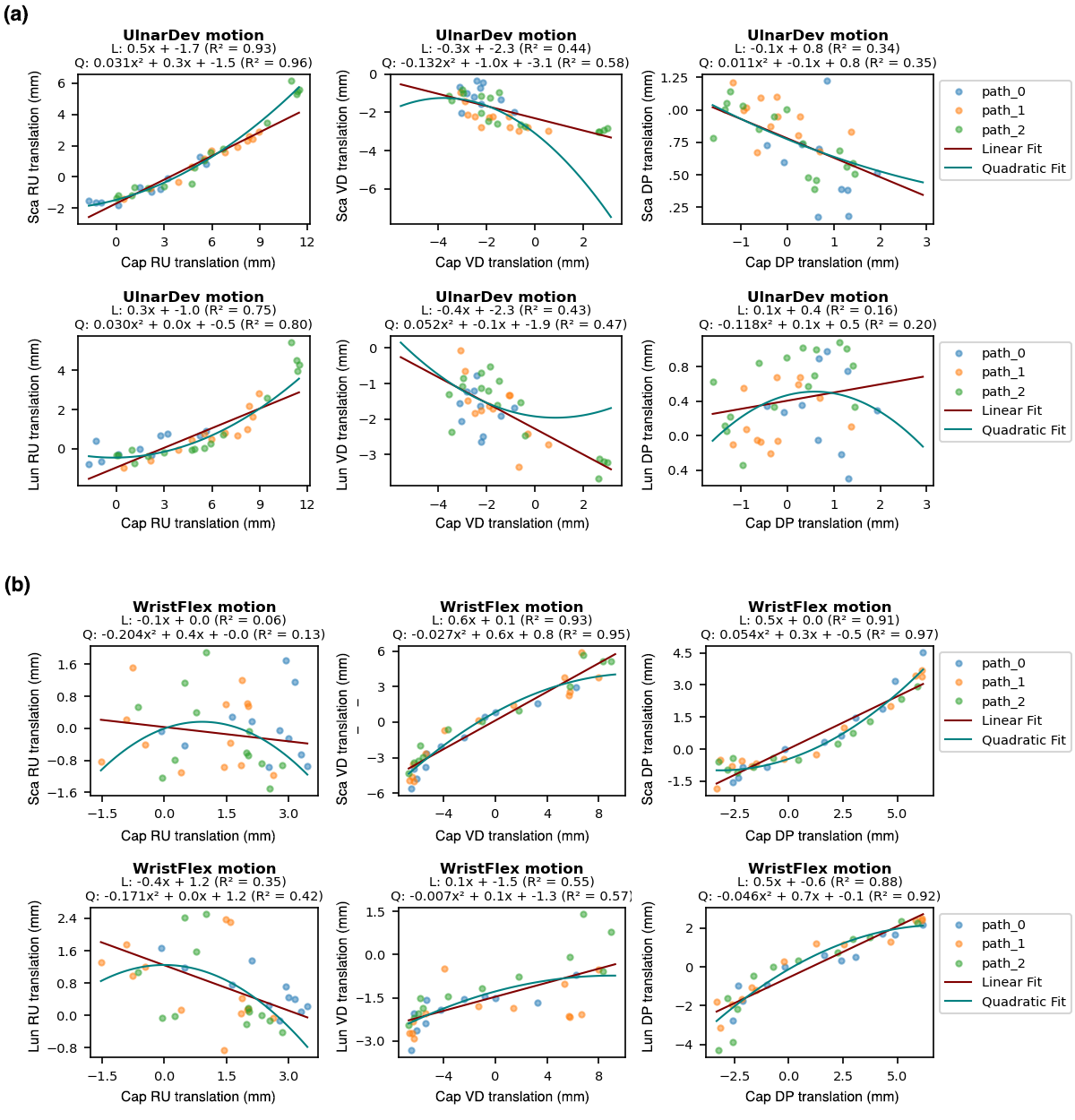}
	\caption{A representative depiction of linear and quadratic models are fitted to scaphoid and lunate translation versus the capitate angle, during (a) radial-ulnar deviation and (b) flexion-extension motion. The fits displayed in the figure represent the average fit calculated across three different paths. RU: radial-ulnar, VD: volar-dorsal, DP: distal-proximal }
	\label{fig:fig6_trans_fit}
\end{figure}

Example fits are displayed in Figures \ref{fig:fig5_ang_fit} and \ref{fig:fig6_trans_fit}.   Since the subjects were unconstrained in their motion, there is an expected variation in the color-coded path trajectories, which can be seen in these sample plots for a single subject. 

\subsection{Kinematic Metric Stability Analysis}

The evaluative focus of the current study was to assess the stability of kinematic metrics derived from the multiple motion cycles conducted by all participating subjects. This assessment was further segregated between two distinct cohorts: asymptomatic and symptomatic subjects. The primary goal of this segmentation was to discern differences in metric stability patterns between the two groups.

In order to quantify reliability of derived metrics, Intraclass Correlation Coefficients (ICC), a statistical measure well-suited to gauge consistency and absolute agreement of metrics within and between groups, were computed using the Python package Pingouin.

For intra-subject stability evaluation, ICC was computed using the ICC(3) model available in the Pingouin package. This model was chosen due to its particular appropriateness for scenarios where the raters, or in this case, the metrics derived from each of the subject's three motion cycles, are considered fixed entities. This allows the ICC to reflect not only the degree of consistency but also the absolute agreement of metrics derived from multiple motion cycles for each individual.

On the other hand, for inter-subject stability, ICC was computed separately for the asymptomatic (n=29) and symptomatic (n=20) cohorts, utilizing Pingouin's ICC1 model. This model was specifically chosen as it measures the absolute agreement among derived metrics across subjects. 

 The distributions of ICC values within each cohort were used to compute a set of basic characteristics representing spread (standard deviation), skewness (symmetry), and central tendency (median). 

To further assess the distribution of ICC values within the two cohorts, rolling distribution percentiles at 90th, 80th, 70th, 60th, and 50th percentile levels. These percentiles furnished a more granular understanding of the distribution and allowed for detailed comparison of stability across the two cohorts.

In addition to examining the overall distribution of ICC values within each cohort, we also performed a differential analysis of ICC values for each metric across the asymptomatic and symptomatic cohorts. This process was instrumental in identifying specific metrics that exhibited a higher degree of variability within the symptomatic cohort as compared to the asymptomatic one. This differential analysis thereby elucidated the unique impact of wrist symptomatology on the variability of these kinematic metrics.

%The script performs an analysis of motion profiles, specifically ulnar deviation and wrist flexion motions, of the scaphoid and lunate bones. 

%Intraclass correlation (ICC) coefficient calculations were derived to analyse reproducibility of the derived metrics within the  for specific metrics derived from these motion profile.

%The ICC measures the degree of consistency and absolute agreement among these trials, giving a statistical measure of the reliability of the metrics.

%A subset of the data corresponding to each combination is selected, and the ICC is calculated using the $intraclass_corr$ function from the pingouin library. This subset is reshaped using the melt function from pandas library, which reshapes the data to a format suitable for the ICC calculation.

\section{Results}

\begin{figure}[h!]
	\centering
	\includegraphics[width=6.5in]{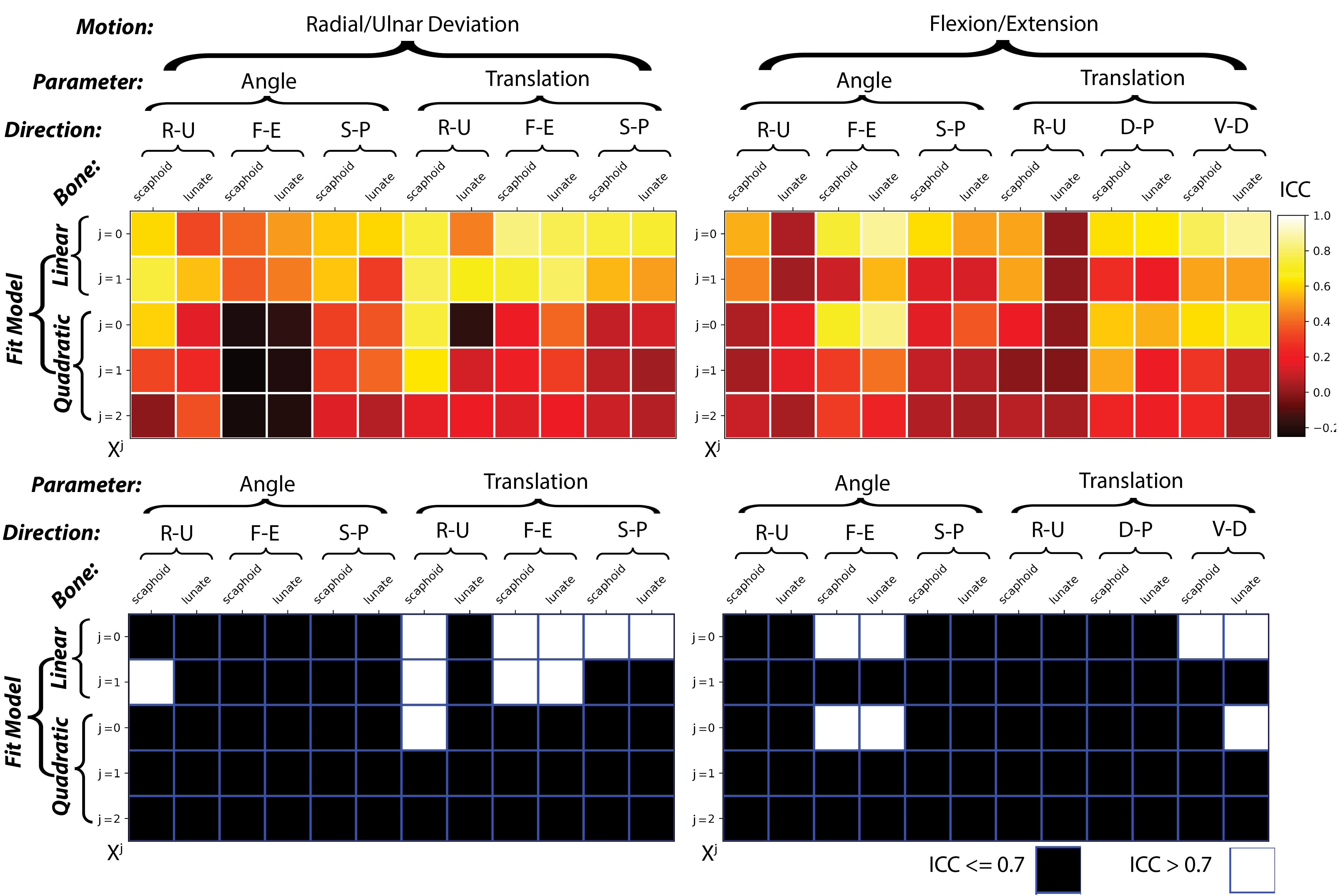}
	\caption{Metric panel maps displaying stability through ICC metrics for intra-subject (assessing the multiple trials) across the full study cohort (both symptomatic and asymptomatic, n=49).   The bottom plots identify those metrics with high ICC, suggesting these are the most stable metrics of the 120 element comprehensive panel.}
	\label{fig:intra-Stability}
\end{figure}

Figure~\ref{fig:intra-Stability} provides a graphical representation of the metric stability analysis performed within the full study cohort.   The figure presents a comprehensive analysis of different metrics, represented in a map-like layout.   In assessing intra-subject stability across the two distinct movements - deviation and flexion/extension - we observed a comparable degree of stability. Out of the total 120 metrics evaluated, 10 metrics achieved an Intraclass Correlation Coefficient (ICC) exceeding 0.7 in the deviation movement, signifying a high level of stability. Conversely, in the flexion/extension movement, 7 metrics reached this level of stability. Interestingly, the highly stable metrics varied between the two motions, indicating that different metrics may provide stability for different types of movement.

When analyzing the types of metrics that exhibited high stability in each movement, a pattern emerged. In the deviation movement, the metrics demonstrating high stability were predominantly centered on translation metrics. This suggests that these metrics might play a significant role in maintaining stability during this specific movement.

However, in the flexion-extension movement, the stable metrics were spread across both rotational changes in flexion-extension and translational changes in the volar-dorsal direction. This indicates a more diverse distribution of stable metrics in this movement, potentially reflecting the complex nature of the flexion-extension movement involving both rotation and translation.

Interestingly, out of all the stable metrics, only one metric, specifically the slope of the scaphoid radial-ulnar rotation angle, was not a zeroth-order term of the polynomial fits. This might suggest that most stable metrics are those that directly describe the position or orientation, rather than changes in these parameters.

Furthermore, the majority of the stable metrics, specifically 13 out of 17, were derived from linear fit models. This dominance of linear models among stable metrics implies that simple linear models may be adequate to describe these metrics' behavior, despite the complex nature of the movements.

\begin{figure}[h!]
	\centering
	\includegraphics[width=6.5in]{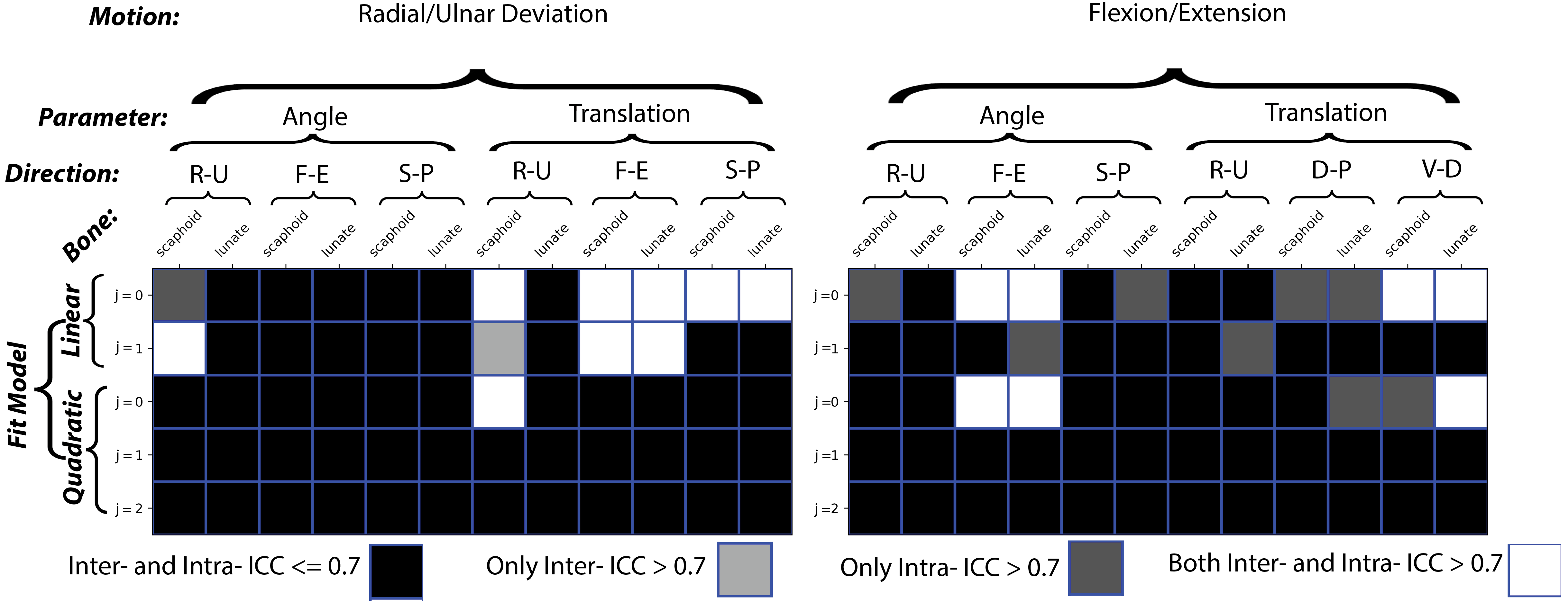}
	\caption{Metric panel map indicating metrics that have ICC <= 0.7 for both inter- and intra- subject analyses, along with overlap across the intra- and inter- subject high stability (ICC>0.7) threshold.  16 of the 17 metrics with high intra-subject ICC also showed high stability in the inter-subject analysis.   Several metrics uniquely showed higher stability in the inter-subject analysis, particularly in the flexion/extension motions.  }
	\label{fig:intrainteroverlap}
\end{figure}

Figure~\ref{fig:intrainteroverlap} analyzes the differences between the ICC values computed in the full-cohort intra-subject analysis and the asymptomatic-only inter-subject analysis.    Of notable interest in the results is the remarkable consistency between the asymptomatic inter-subject and intra-subject analyses. Out of 17 metrics identified with high intra-subject ICC, 16 of them also exhibited high stability in the inter-subject analysis. This level of congruence suggests that these metrics, despite the natural variations across subjects, demonstrate consistent outcomes within each subject as well.

However, the data also revealed interesting differences. Some metrics showed higher stability exclusively in the inter-subject analysis, suggesting that these metrics may be more robust to individual differences and potentially more useful in comparative studies across subjects. This phenomenon was particularly notable in metrics related to flexion/extension motions. Such metrics might, therefore, be more reliable for comparative studies across a diverse range of subjects.

\begin{figure}[h!]
	\centering
	\includegraphics[width=4.0in]{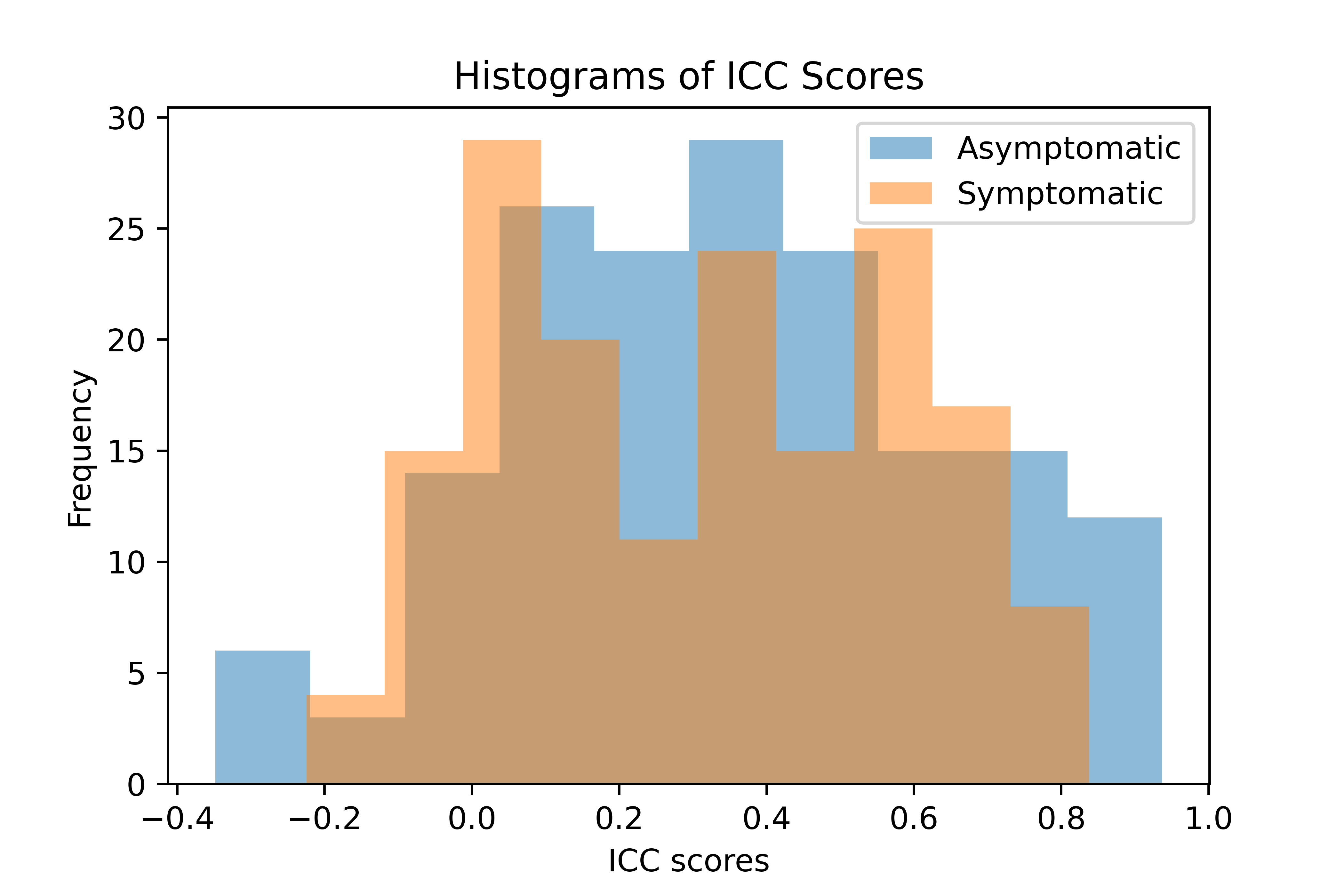}
	\caption{Histograms between cohorts}
	\label{fig:inter-hist}
\end{figure}

% \begin{figure}[h!]
% 	\centering
% 	\includegraphics[width=4.0in]{AvsSbarPlots.png}
% 	\caption{Counts}
% 	\label{fig:inter-counts}
% \end{figure}

\begin{figure}[h!]
	\centering
	\includegraphics[width=6.5in]{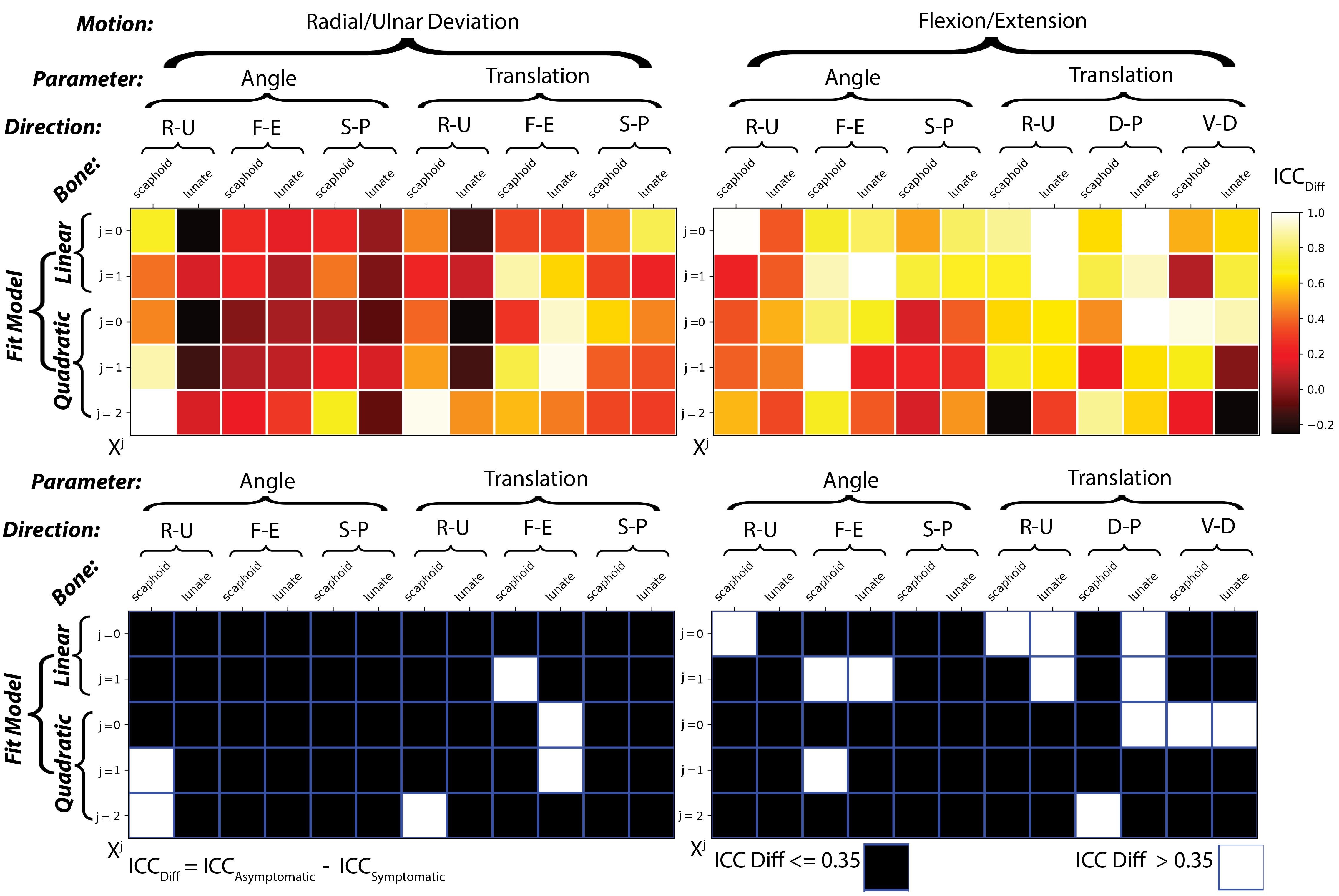}
	\caption{Difference in Inter-Subject stability between asymptomatic and symptomatic cohorts.  Heat maps of ICC differences between the cohort ICC computations are displayed (positive difference indicated higher stability in the asympomatic cohort).  Specific metrics showing higher stability in the asymptomatic cohort are indicated in the binary maps.  }
	\label{fig:inter-diff}
\end{figure}

Adding the ICC analysis within the symptomatic-only cohort offers insight into the variance of computed metrics as a function of previous wrist injury. 
When separately computed within the asymptomatic and symptomatic cohorts, ICC values showed notable differences in distribution trends and specific metrics.  For visualization purposes, histograms of the two ICC distributions are shown in Fig.~\ref{fig:inter-hist}.

Comparative analysis of these two distributions are presented in Table~\ref{tab:inter-stability}. The statistics computed include various percentiles, standard deviation, skewness, and the median.

\begin{table}[h!]
\centering
\begin{tabular}{lrr}
\toprule
Statistic & Asymptomatic & Symptomatic \\
\midrule
90th Percentile & 0.76 & 0.68 \\
80th Percentile & 0.61 & 0.57 \\
70th Percentile & 0.48 & 0.52 \\
60th Percentile & 0.41 & 0.40 \\
50th Percentile & 0.36 & 0.33 \\
Standard Deviation & 0.29 & 0.27 \\
Skewness & -0.01 & 0.06 \\
Median & 0.36 & 0.33 \\
\bottomrule
\end{tabular}
\caption{Summary results from comparative analysis of ICC values computed on distributions of metrics within the asymptomatic and symptomatic cohorts. }
\label{tab:inter-stability}
\end{table}

In the 90th and 80th percentile, the ICC values were higher in the asymptomatic group, indicating a higher degree of reliability at these percentiles in the asymptomatic group compared to the symptomatic group. However, at the 70th percentile, the ICC values in the symptomatic group exceed those in the asymptomatic group. For the 60th and 50th percentiles, ICC values remain higher in the asymptomatic group, albeit the differences are less pronounced.

The standard deviation of ICC values in the asymptomatic group is slightly higher than that in the symptomatic group, suggesting a wider spread in the ICC values among the asymptomatic subjects.

Skewness is close to zero in both groups, but with a slight negative skewness in the asymptomatic group and a slight positive skewness in the symptomatic group. This suggests that the ICC distribution is slightly left-skewed in the asymptomatic group and slightly right-skewed in the symptomatic group, but the skewness is relatively minor in both cases, suggesting the distributions are close to symmetrical.

The median ICC value is slightly higher in the asymptomatic group than in the symptomatic group, indicating that the central tendency of the ICC distribution is higher among asymptomatic subjects.

Overall, this distribution analysis of ICC values suggests some variability in the reliability of the assessed kinematic metrics between asymptomatic and symptomatic subject cohorts, with generally higher reliability in the asymptomatic group, particularly in the higher percentiles.

The number of derived metrics that showed high levels of stability (ICC > 0.7) was higher in the asymptomatic (25) vs symptomatic cohorts (12).

% Figure~\ref{} provides graphical analyses of differentiating the iinter-Subject stability between asymptomatic and symptomatic cohorts.  Heat maps of ICC differences between the cohort ICC computations are displayed (positive difference indicated higher stability in the asympomatic cohort).  Specific metrics showing higher stability in the asymptomatic cohort are indicated in the binary maps.  Of note, there are notable increases in stability within the asympomatic cohort in quadratic fit terms that did not show substantial initial stability in the intra-subject or asymptomatic-only ICC analyses.   This suggests that quadratic term metrics may have substantial variances in symptomatic cohorts that may be of use, despite their relatively lower levels of stability within normative subjects.    

Figure~\ref{fig:inter-diff} offers a visual comparison of inter-subject stability within the asymptomatic and symptomatic cohorts. 

Heat maps are again utilized for visualization purposes, though the maps now reflect the differences in ICC between the two cohorts. A positive difference indicates higher stability within the asymptomatic cohort, illuminating how certain metrics may be more consistent in asymptomatic subjects.

Binary maps are also again employed to spotlight specific metrics that show higher stability in the asymptomatic cohort. 

These cross-cohort analyses reveal a noticeable increase in stability within the asymptomatic cohort concerning the quadratic fit terms. This increase is substantial, even though these terms did not demonstrate considerable initial stability in either the intra-subject analysis or the asymptomatic-only ICC analysis. This observation suggests that the quadratic term metrics may possess significant variances in symptomatic cohorts.   Despite their relative instability within normative (asymptomatic) subjects, these quadratic term metrics could potentially serve as valuable indicators of disease symptoms or severity.

\section{Discussion and Conclusion}
In this study, we have outlined the construction of a scaphoid and luate kinematic metric panel using the 4D dynamic wrist MRI methodology recently published by Zarenia et al~\cite{zarenia2022dynamic}.   The demonstrated kinematic metric panel consists of 120 metrics per subject, which are computed by fitting low order polynomial models of scaphoid and lunate degrees of freedom against the capitate for two types of unconstrained wrist movement -- radial/ulnar deviation and flexion extension.

On a mixed cohort of 49 subjects with and without a history of wrist injury, we evaluated the stability of these metrics through the computation of intraclass correlation coefficients (ICC).   ICC values were computed within the context of both intra- (within subject) and inter- (across subject) metric stability. 

Our analysis revealed noteworthy insights into intra-subject stability of the derived carpal kinematic metrics.  For each subject, 3 metric values were determined for each motion using automated detection of the 3 motion cycles each subject performed.   Comparable degrees of stability were observed between the radial/ulnar deviation and flexion/extension wrist motions, with 10 and 7 metrics respectively achieving an ICC exceeding 0.7. It is interesting to note that the specific metrics demonstrating high stability varied between the two motions, pointing to the possibility that different metrics are key to stability depending on the type of movement.

A closer look at the nature of these stable metrics revealed patterns specific to each type of movement. In the case of the deviation movement, translation metrics were predominantly associated with high stability. However, the flexion-extension movement exhibited a broader range of stable metrics, spanning both rotational and translational changes.   Metrics derived from linear, rather than quadratic fit models, dominated the list of most stable (ICC > 0.7) intra-subject stable metrics, with 13 out of 17 coming from this category. 

The stability of the derived kinematic metrics was further examined in an inter-subject analysis of asymptomatic individuals. Interestingly, out of the 17 metrics with high intra-subject ICC, 16 also showed high inter-subject stability in the asymptomatic cohort. This high level of agreement suggests that these metrics are not only consistent within individuals but also across different asymptomatic individuals, reinforcing their potential utility in future analyses of carpal kinematics.

A comparative analysis of asymptomatic and symptomatic cohorts revealed a higher number of stable metrics in the asymptomatic group. Interestingly, despite their relative instability in intra-subject analyses, quadratic term metrics demonstrated substantial changes in stability between the symptomatic and asymptomatic subjects, suggesting their potential relevance in identifying disease symptoms or severity.

In conclusion, the present study has presented the construction of a diverse set of self-referenced kinematic metrics reflective of scaphoid and lunate motion during unconstrained radial/ulnar deviation and flexion/extension movements captured using a novel 4D-MRI acquisition paradigm.    Within a mixed cohort of subjects with and without wrist injury,  insights into the relative stability of the derived carpal kinematic metrics was evaluated.   The results of this analysis have identified patterns of metric stability within subjects and demonstrated clear differences in stability of select metrics in asymptomatic and symptomatic metrics.   The general agreement between stable metrics in within subjects and across the asymptomatic cohort provides evidence of normative metric consistency.   A relative reduction of stable metrics in the symptomatic subjects provides further encouraging evidence that the derived kinematic metrics hold promise as indicators of carpal dysfunction.   

\section{Acknowledgment}
The author would like to express their gratitude to Volkan Emre Arpinar, PhD, Mohammad Zarenia, PhD, Brad Swearingen, and Robin Ausman for their valuable assistance in this study.

Research reported in this publication was supported by National Institute of Health (NIH) grant R21AR075327.

\bibliographystyle{unsrtnat}
\bibliography{references}  %%% Uncomment this line and comment out the ``thebibliography'' section below to use the external .bib file (using bibtex) .

\end{document}